\documentstyle[12pt]{article}
\newcommand{\be}{\begin{equation}}
\newcommand{\bea}{\begin{eqnarray}}
\newcommand{\eea}{\end{eqnarray}}
\newcommand{\ba}{\begin{array}}
\newcommand{\ea}{\end{array}}
\newcommand{\ee}{\end{equation}}

\expandafter\ifx\csname mathbbm\endcsname\relax

\newcommand{\cn}{{\cal N}}
\else

\fi \textheight 22cm \textwidth 15cm \topmargin 1mm
\def\l{\label}

\begin{document}
\begin{titlepage}
\hfill
\vbox{
    \halign{#\hfil         \cr
           IPM/P-2003/011 \cr
           hep-th/0303074  \cr
           } 
      }  
\vspace*{20mm}
\begin{center}
{\Large {\bf On the Multi Trace Superpotential and Corresponding
Matrix Model}\\ }

\vspace*{15mm}
\vspace*{1mm}
{Mohsen Alishahiha$^a$ \footnote{alishah@theory.ipm.ac.ir}
and Hossein Yavartanoo$^{a,b}$}
\footnote{yavar@theory.ipm.ac.ir} \\
\vspace*{1cm}

{\it$^a$ Institute for Studies in Theoretical Physics
and Mathematics (IPM)\\
P.O. Box 19395-5531, Tehran, Iran \\ \vspace{3mm}
$^b$ Department of Physics, Sharif University of Technology\\
P.O. Box 11365-9161, Tehran, Iran}\\

\vspace*{1cm}
\end{center}

\begin{abstract}
We study ${\cal N}=1$ supersymmetric $U(N)$ gauge theory coupled to 
an adjoint scalar superfiled with a cubic superpotential
containing a multi trace term. We show that the field theory 
results can be reproduced from a matrix model which its 
potential is given in terms of a linearized potential obtained 
from the gauge theory superpotential by adding some auxiliary 
nondynamical field. Once we get the effective action from this 
matrix model we could integrate out the auxiliary field getting 
the correct field theory results.

\end{abstract}

\end{titlepage}

\section{Introduction}

Recently it has been proposed \cite{Dijkgraaf:2002dh} that the 
exact superpotential and gauge coupling for a wide class of 
${\cal N}=1$ supersymmetric gauge theories can be obtained using 
perturbative computations in a {\it related} matrix model.
Given an ${\cal N}=1$ SYM theory the potential of the corresponding
matrix model is given in terms of the gauge theory superpotential. 
Even more interesting the nonperturbative results of gauge theory can be 
obtained from just planar diagrams of matrix model without taking any large 
$N$ limit in the gauge theory. 
This conjectured based on earlier works \cite{Bershadsky:1993cx}
-\cite{Dijkgraaf:2002vw} and has recently been verified perturbatively using 
superspace formalism \cite{Dijkgraaf:2002xd} or anomalies 
\cite{{Gorsky:2002uk},{Cachazo:2002ry}}. Further developments can be found in 
\cite{Seiberg:2002jq}-\cite{Brandhuber:2003va}.

To make the proposal precise and also to fix our notation consider a
$U(N)$ gauge theory with ${\cal N}=1$ supersymmetry coupled to a chiral
superfield in the adjoint representation of $U(N)$. Moreover, in general,
we take the following superpotential
\be
W(\phi)=\sum_{k=1}^{n+1}{g_k\over k}{\rm Tr}(\phi^{k})
\ee
for some $n$. To get a supersymmetric vacuum we need to impose D- and 
F-term conditions. Taking $\phi$ to be diagonal would satisfy the 
D-term and for F-term we need to set $W'(\phi)=0$. This equation has 
$n$ roots $a_i$ and thus one can write $W'(x)=g_{n+1}\prod_{i=1}^n(x-a_i)$. 
Therefore by taking $\phi$ to have eigenvalue $a_i$ with multiplicity 
$N_i$, the gauge symmetry $U(N)$ is broken down to
$\prod_{i=1}^nU(N_i)$ with $\sum_{i=1}^nN_i=N$.

If the roots $a_i$ are all distinct, the chiral superfields are all
massive and can then be integrated out to get an effective action
for low energy theory. The chiral part of the low energy effective 
Lagrangian can be written as \cite{Cachazo:2002ry}
\be
L_{\rm eff}=\int d^2\theta\; W_{\rm eff}+c.c.\;,\;\;\;\;\;\;
W_{\rm eff}=f(S_k,g_k)+\sum_{i,j}\tau_{ij}\;\omega_{\alpha i}
\omega^{\alpha}_j\;,
\ee 
where $S_k=-{1\over 32\pi^2}{\rm Tr}W_{\alpha i}W^{\alpha i}$ and
$\omega_{\alpha i}={1\over 4\pi}{\rm Tr}W_{\alpha i}$ with 
$W^{\alpha i}$ being the gauge superfields of $U(N_i)$ gauge group.

The main point in the Dijkgraaf-Vafa's proposal is that the
chiral part of the effective action can be given by a holomorphic
function ${\cal F}_G(S_k)$, such that
\be
W_{\rm eff}=\sum_{i=1}^nN_i{\partial {\cal F}_G\over \partial S_i}
+{1\over 2}\sum_{i,j=1}^n 
{\partial^2 {\cal F}_G\over \partial S_i\partial S_j}\omega_{\alpha i}
\omega^{\alpha}_j\;.
\ee
Now what is left to be determined is the function ${\cal F}_G$. In fact it 
is the goal of the Dijkgraaf-Vafa's proposal to identify ${\cal F}_G$ as the
free energy of an auxiliary nonsupersymmetric matrix model which has for
its ordinary potential  the same function $W$ that is the superpotential
of the four dimensional supersymmetric gauge theory. The matrix model free 
energy is given by
\be
e^{{1\over g_s^2}{\cal F}_0}={1\over {\rm Vol}(U(M))}
\int D\phi\; e^{\left(-{1\over g_s}W(\phi)\right)}\;,
\l{FREE}
\ee
where $\phi$ is an $M\times M$ matrix belongs to $U(M)$. For the model
we are considering one needs also to take $\phi$ in such a way that the 
$U(M)$ symmetry is broken down to $\prod_{i=1}^nU(M_i)$ such that
$\sum_{i=1}^nM_i=M$. Moreover one should also identify $S_i=g_sM_i$.
Taking large $M$ limit one can compute ${\cal F}_0$ 
order by order using only planar diagrams in the matrix perturbation theory. 
Now the prescription \cite{Dijkgraaf:2002xd} is that, for example, the 
$l$th instanton contribution to the effective action can be reproduced
from a perturbative contribution with $l$ loops in the auxiliary matrix
model. Actually having the matrix model free energy the effective
superpotential is obtained by
\be
W_{\rm eff}=\sum_{i=1}^n\left(N_i{\partial {\cal F}_0\over \partial S_i}
-2\pi i\tau_0 S_i\right)\;.
\l{lll}
\ee
where $\tau_0$ is the bare coupling of the theory.  

By now there are huge number of papers devoted to this proposal where
only superpotentials with single trace operators have been studied. 
Recently superpotentials containing multi trace operators has also been
considered in \cite{Balasubramanian:2002tm} where the authors showed that
taking naively $W$ with multi trace as the potential of matrix model 
would lead to {\it incorrect} matrix model. By ``incorrect'' they mean 
that one cannot reproduce the corresponding gauge theory results,
though the obtained matrix model could be an auxiliary matrix model of
some gauge theory which, of course, is not what we started with.
More precisely it has been shown that although the diagrams surviving
the large $M$ limit of the matrix model with multi trace potential 
are exactly the graphs that contribute to the effective action
of the field theory with multi trace tree level superpotential, one
cannot compute the effective superpotential of the field theory
by taking a derivative ${\partial {\cal F}_0\over \partial S}$. 

This problem can, of course, be solved \cite{Balasubramanian:2002tm}
using the linearized superpotential in the matrix model. In fact
starting with a multi trace operators in the superpotential one can 
linearized it using some nondynamical background fields $A_i$. Then
the potential would contain only single trace operators with 
$A_i$'s dependent coefficients. Once we find $W_{\rm eff}$ from matrix
model, we can integrate out $A_i$'s fields getting the {\it correct}
gauge theory result. 

This is the aim of this article to further study a superpotential
containing  multi trace operators. In fact we shall study ${\cal N}=1$ 
$U(N)$ SYM theory coupled to an adjoint scalar superfield with the cubic 
superpotential given by \footnote{${\cal N}=1$ supersymmetric $U(N)$
gauge theory
with cubic single trace superpotential has been extensively studied, 
for example, in \cite{Dijkgraaf:2002pp}-\cite{Dijkgraaf:2002yn}.}
\be
W_{\rm tree}={1\over 3}\;{\rm Tr}(\phi^3)+{1\over 2}\;m\;{\rm
Tr}(\phi^2)+\lambda
\;{\rm Tr}(\phi)+{1\over 2}\;g\;{\rm Tr}(\phi)\; {\rm Tr}(\phi^2)\;.
\ee
We will see that, using the linearized form of the superpotential, one 
can reproduce the gauge theory results in the cases with and without
gauge symmetry breaking. We note that in \cite{Balasubramanian:2002tm}
the authors have only considered a model where the gauge symmetry is
not broken. As we will see the procedure works in the case with 
broken gauge symmetry as well.

The organization of this paper is as follows. In section 2 we will review
${\cal N}=1$ $U(N)$ SYM theory with cubic single trace superpotential.
We shall consider two different cases in which the gauge group may or may 
not be broken. In section
3 we will study the same theory with a multi trace term added to the
superpotential. Regarding the fact that this model can be though of 
as a deformation of ${\cal N}=2$ theory we will find the effective
superpotential using the factorization of Seiberg-Witten curve. In 
section 4 we will reproduce the same field theory results using
linearized superpotential. In section 5 we will see how the corresponding
matrix model can be treated. The last section is devoted to conclusions. 
Some technical computation of factorization of Seiberg-Witten curve is
presented in the appendix.

\section{Single trace superpotential}

In this section we shall review the ${\cal N}=1$ supersymmetric $U(N)$
gauge theory coupled with an adjoint scalar hypermultiplet with
cubic superpotential containing only single trace operators
\be
W_{\rm tree}={1\over 3}\;{\rm Tr}(\phi^3)+{1\over 2}\;m\;
{\rm Tr}(\phi^2)+\lambda\;
{\rm Tr}(\phi)\;.
\l{WT}
\ee
Taking $\phi$ diagonal one just needs to set $W'(\phi)=0$ to get the 
supersymmetric vacuum, and therefore the derivative of 
superpotential can be recast to
\be
W'(x)=(x-a_1)(x-a_2),\;\;\;\;\;a_{1,2}=-{m\over 2}\pm
{1\over 2}\sqrt{m^2-4\lambda}\;.
\ee
In general we can take $\phi$ to have eigenvalues $a_1$ or $a_2$ with
multiplicity $N_1$ and $N_2$, respectively. This will break gauge symmetry 
to $U(N_1)\times U(N_2)$ with $N_1+N_2=N$. Of course as an special case
one can, for example, take $N_2=0$ which corresponds to the supersymmetric
vacuum without gauge symmetry breaking. In the 
following we shall consider both cases.

\subsection{Unbroken gauge symmetry}

In this section for the case where the gauge symmetry is not broken, 
we will first review how the exact superpotential can be obtained using 
the factorization of the Seiberg-Witten curve. In fact the model we are 
interested in can be obtained from ${\cal N}=2$ supersymmetric
$U(N)$ gauge theory perturbed by a general tree level superpotential 
given by
\be
W_{\rm tree}=\sum_{k=1}^{n+1}{1\over k}g_k{\rm Tr}(\phi^k)\;.
\l{WGEN}
\ee
A generic point in the moduli space of the  $U(N)$  ${\cal N}=2$ 
theory will be lifted by adding such a superpotential.  
The points which are not lifted
are precisely where at least $N-n$ mutually local monopoles 
become massless. This can be seen from the following argument.
The gauge group in the ${\cal N}=1$ theory is broken down to 
$\prod_{i=1}^n U(N_i)$, and each $SU(N_i)$ factors are confined. 
We expect  condensation of $N_i -1$ 
magnetic monopoles in each of  these $SU(N_i)$ factors and a total of $N-n
$ condensed magnetic monopoles.  These monopoles condense at the points on the 
${\cal N}=2$ moduli space where $N-n$ mutually local monopoles become massless. 
These are  precisely the points which are not lifted by addition of the
superpotential. 

These considerations are equivalent to the requirement that the corresponding 
Seiberg-Witten curve has the factorization
\begin{equation}
P_N^2(x,u)-4 \Lambda^{2N}=H_{N-n}^2(x) F_{2n}(x),
\end{equation}
where $P_N(x,u)$ is an order $N$ polynomial in $x$ with coefficients
determined by the (vevs of) the $u_k$, $\Lambda$ is an ultraviolet cut-off,
$H$ and $F$ are order $N-n$ and $2n$ polynomials in $x$, respectively.

The $N-n$ double roots place $N-n$ conditions on the original variables
$u_k$. We can parameterize all the $\langle u_k\rangle $ by $n$ independent
variables $\alpha_j$.  In other words, the $\alpha_j$'s then correspond
to massless fields in the low-energy effective theory.  If we know
the exact effective action for these fields, to find the vacua, we simply
minimize $S_{eff}$. Furthermore, substituting $\langle u_k \rangle$ back
into the effective action gives the action for the vacua.

In general the factorization problem is hard to solve,
but for the confining vacuum where all $N-1$ monopoles have condensed,
there is a general solution given by Chebyshev polynomials.\footnote{This
was worked out first by Douglas and Shenker \cite{Douglas:1995nw}.} 
In our case, we have the solution
\newcommand{\Cpq}[2]{{\left( \begin{array}{c} {#1} \\ {#2} \end{array}
\right)}}
\be
\label{confsol}
\langle u_p\rangle={N\over p} \sum_{q=0}^{[p/2]} C_{p}^{2q} C_{2q}^q 
\Lambda^{2q}
z^{p-2q},\;\;\;\;\;C_n^p := \Cpq{n}{p} = \frac{n!}{p!(n-p)!}\;, 
\ee
where $z=\frac{\langle u_1\rangle}{N}$.
We note, however, that the above procedure  
is not the best form to be
compared with the matrix model result because there is no gluino 
condensate
$S$. To make the comparison, we need to ``integrate in'' 
\cite{Intriligator:1994uk} the glueball superfield as in 
\cite{Ferrari:2002jp}.

In the model we are considering the exact
superpotential which has to be minimized is
\be
W_{\rm exact}=\langle u_3\rangle +m \langle u_2\rangle +\lambda 
\langle u_1\rangle\;,
\l{WTE}
\ee
where
\be
\langle u_1\rangle=Nz,\;\;\;\;\;\langle u_2\rangle={N\over 2}(z^2+2\Lambda^2),
\;\;\;\;\;\langle u_3\rangle={N\over 3}(z^3+6\Lambda^2z)\;.
\ee

The ``integrate in'' procedure can be done by setting 
$B := \Lambda^2$, and use the equation 
\be
N S=B {\partial W_{\rm exact}\over \partial B}=NB(m+2z)\;,
\l{Z1}
\ee
to solve for $B$ in terms of $S$. Then we find $z$ by solving
\be
0={\partial W_{\rm exact}\over \partial z}=N(z^2+mz+\lambda+2B)\;.
\l{Z2}
\ee
Now the effective superpotential for the glueball superfield $S$ can 
be written 
as
\be
W_{\rm eff}(S,g,\Lambda)= -S \log\left({B \over \Lambda^2}\right)^N +
W_{\rm exact}(S, g)\;.
\ee
One can explicitly solve the equations (\ref{Z1}) and (\ref{Z2}) to 
find $z$ and $B$ in 
power series of $S$. The results up to ${\cal O}(S^6)$ are
\bea
B&=&{S\over \Delta}+4{S^2\over \Delta^4}+40{S^3\over \Delta^7}
+512{S^4\over \Delta^{10}}+7392{S^5\over \Delta^{13}}\;,\cr &&\cr
z&=&{-m+\Delta\over 2}-2{S\over \Delta^2}-12{S^2\over \Delta^5}
-128{S^3\over \Delta^8}
-1680{S^4\over \Delta^{11}}-24576{S^5\over \Delta^{14}}\;.
\eea
Plugging these solutions to the above expression of effective
superpotential, one gets
\be
W_{\rm eff}=-NS(\log({S\over \Delta \Lambda^2})-1)
-{2N\over 3}\;{S^2\over \Delta^3}\left(3+16{S\over \Delta^3}+
140{S^2\over \Delta^6}+512{S^3\over \Delta^9}\right)\;,
\l{hhhh}
\ee
which is the exact effective action up to 5 instanton. 

Using the Dijkgraaf-Vafa's proposal one will be able to reproduce this
result using a nonsupersymmetric matrix model with the potential
given by (\ref{WT}). Since we are interested in the case where the 
gauge symmetry is not broken one considers the expansion around a 
classical solution as the following
\be
\phi=a_1 1_{M\times M}+\varphi\;,
\ee
and therefore the potential of matrix model reads
\be
W(\varphi)=W(a_1)+{1\over 3}\;{\rm Tr}(\varphi^3)+{1\over 2}\;\Delta\;
{\rm Tr}(\varphi^2)\;,
\ee
where $\Delta=a_1-a_2$. Here $\phi$ is $M\times M$ matrix belongs to 
$U(M)$ group. One can now write the Feynman rules and thereby 
evaluate the matrix model free energy order by order using 
(\ref{FREE}). Here we shall take a limit in which $M$ is large and keeping 
the 't Hooft coupling $S=g_sM$ fixed, and thus only planar diagrams would
contribute. 
In this limit the free energy is found to be \cite{Dijkgraaf:2002pp}
\be
{\cal F}_0(S)={1\over 2}S^2\log\left({S\over \Delta^3}\right)-S^2
\log\left({\Lambda\over \Delta}\right)+{2\over 3}\;{S^3\over\Delta^3}
\left(1+4{S\over \Delta^3}+28{S^2\over \Delta^6}+\cdots\right)
\ee
up to 4-loop. Using this expression the exact superpotential is given by
(see also \cite{Cachazo:2001jy})
\bea
W_{\rm exact}=-NS\left(\log\left({S\over \Delta\Lambda^2}\right)-1\right)-
{2N\over 3}\;{S^2\over\Delta^3}
\left(3+16{S\over \Delta^3}+140{S^2\over \Delta^6}+\cdots\right)\;,
\l{WEXACT}
\eea
which is in exact agreement with the field theory computation
(\ref{hhhh}). As we see the $l$th loop contribution to the
matrix model free energy is the same as $l$ instantons contribution
to the effective action.

\subsection{Broken gauge symmetry}

In this subsection we shall review the case where the gauge symmetry 
is broken to two parts. In other words we consider a matrix model where 
$U(M)$ group is broken down to $U(M_1)\times U(M_2)$. To get such a 
matrix model we take
\be
\phi=\pmatrix{a_11_{M_1\times M_1} & 0\cr 0 & a_21_{M_2\times M_2}}+
\pmatrix{\varphi_{11}&\varphi_{12}\cr \varphi_{21}&\varphi_{22}}
\ee 
here $M_1+M_2=M$. Moreover we will consider the large $M_1$ and $M_2$ 
limit while keeping $S_1=g_sM_1$ and $S_2=g_sM_2$ fixed. 
These means that only planar diagrams would be important. To do the 
explicit computation one can use a gauge in which $\varphi_{12}$ and 
$\varphi_{21}$ are set to zero. This can be done using Faddeev-Popov 
ghost field, and therefore the matrix model action is found as 
\cite{Dijkgraaf:2002pp}
\bea
W&=&{1 \over 2}\;\Delta\bigg{(}{\rm Tr}(\varphi_{11}^2)-{\rm Tr}(\varphi_{22}^2)
\bigg{)}+{1\over 3}\bigg{(}{\rm Tr}(\varphi_{11}^3)+{\rm Tr}(\varphi_{22}^3)
\bigg{)}\cr &&\cr
&+&\Delta\bigg{(}{\rm Tr}(B_{21}C_{12})-{\rm Tr}(B_{12}C_{21})\bigg{)}+
{\rm Tr}(B_{21}\varphi_{11}C_{12}+C_{21}\varphi_{11}B_{12})\cr &&\cr
&+&{\rm Tr}(B_{12}\varphi_{22}C_{21}+C_{11}\varphi_{22}B_{21})\;,
\eea
where $B$ and $C$ are corresponding ghost fields.

It is now easy to write down the Feymann rules for double line Feymann
diagrams and thereby to compute the matrix model free energy order by order.
The result up to 4-loop is \cite{Dijkgraaf:2002pp}
\bea
&&{\cal F}_{0}(S_1,S_2)=-{1\over 2}\sum {S_i}^2\log\left({S_i\over \Delta^3}
\right)+(S_1+S_2)^2 \log\left({\Lambda\over \Delta^3}\right)\cr &&\cr
&&\;\;\;\;+{1\over 3\Delta^3}
\bigg{(}2S_1^3-15S_1^2S_2+15S_1S_2^2-2S_2^3\bigg{)}\cr &&\cr
&&\;\;\;\;+{1\over 3\Delta^6}
\bigg{(}8S_1^4-91S_1^3S_2+177S_1^2S_2^2-91S_1S_2^3-8S_2^4\bigg{)}
\cr &&\cr
&&\;\;\;\;+{1\over 3\Delta^9}
\bigg{(}56S_1^5-871S_1^4S_2+2636S_1^3S_2^2-2636S_1^2S_2^3
+871S_1S_2^4-56S_2^5\bigg{)}.
\eea
Having the matrix model free energy the effective
superpotential for the case where the gauge symmetry is broken 
as $U(N)\rightarrow U(N_1)\times U(N_2)$ can be found as following
\be
W_{\rm eff}=\sum_{i=1}^2\left(N_i{\partial {\cal F}_0\over \partial S_i}
-2\pi i\tau_0 S_i\right)\;.
\ee
where $\tau_0$ is the bare coupling of the theory.

This effective superpotential should be compared with that obtained from
gauge theory computation. The gauge theory result may be found using
factorization of Seiberg-Witten curve, though, in general the factorization
procedure is difficult to be done. Nevertheless for an special case this can 
easily be worked out. For example consider the SYM theory with gauge
group $U(3N)$ broken down into $U(2N)\times U(N)$. Actually the analysis of this
theory is equivalent to SYM theory with gauge group $U(3)$ broken down into
$U(2)\times U(1)$ where the effective superpotential is turned out to be
\footnote{In appendix A we have presented the factorization of
Seiberg-Witten curve 
for the general case where the unbroken gauge symmetry has only one nonabilian
factor of $U(2)$.}\cite{Elitzur:1996gk}
\be
W_{\rm eff}=u_3+m u_2+\lambda u_1\pm 2\Lambda^3\;.
\l{WESU2}
\ee
Of course it is not a suitable form for comparison with the matrix 
model result. Actually to compare these two results one can, for example,
integrate out the $S_1$ and $S_2$ fields from the effective superpotential
obtained from the matrix model. Doing so, one can see that
the matrix model reproduce the correct result (\ref{WESU2}) order by
order \cite{Cachazo:2001jy}.

\section{Multi trace superpotential}

In this section we will study ${\cal N}=1$ $U(N)$ SYM theory coupled to
an adjoint
scalar superfield with a superpotential containing a multi trace operator 
\be
W_{\rm tree}={1\over 3}\;{\rm Tr}(\phi^3)+{1\over 2}\;m\;
{\rm Tr}(\phi^2)+\lambda\;{\rm Tr}(\phi)+{1\over 2}\;g\;
{\rm Tr}(\phi)\; {\rm Tr}(\phi^2)\;.
\l{WTM}
\ee
To get the supersymmetric vacuum one needs to impose F- and D-terms conditions.
Taking diagonal $\phi$ would satisfy the D-term condition and for F-term
we need to solve $W'_{\rm tree}(\phi)=0$. This equation has two solutions
, $b_{1,2}$, and 
therefore in general $\phi$ can be taken such that to have eigenvalue
$b_i$ with multiplicity $N_i$. This will break the gauge symmetry down
to $U(N_1)\times U(N_2)$ with $N_1+N_2=N$. 

To find the eigenvalues $b_i$'s we note that 
the adjoint scalar has been taken as $\phi={\rm diag}(b_1
1_{N_1\times N_1}, b_2 1_{N_2\times N_2})$, and thus the 
superpotential is given by
\bea
W_{\rm tree}&=&
{1\over 3}(N_1b_1^3+N_2b_2^3)+{m\over 2}(N_1b_1^2+N_2b_2^2)+\lambda
(N_1b_1+N_2b_2)\cr &&\cr &+&{g\over 2}(N_1b_1+N_2b_2)(N_1b_1^2+N_2b_2^2)\;,
\eea
so, the F-term condition reads
\bea
\lambda+mb_1+b_1^2+{g\over 2}(N_1b_1^2+N_2b_2^2)+gb_1(N_1b_1+N_2b_2)&=&0\cr
\lambda+mb_2+b_2^2+{g\over 2}(N_1b_1^2+N_2b_2^2)+gb_2(N_1b_1+N_2b_2)&=&0\;.
\eea
One can now solve these equations to find $b_1$ and $b_2$. The solution is
\be
b_1=-{m\over 1+N_1g}-{1+N_2g\over 1+N_1g}\;b_2,
\ee
and $b_2$ satisfies
\be
b_2^2+{\tilde m}b_2+{\tilde \lambda}=0\;,
\ee
where 
\bea
{\tilde m}&=&\frac{(1+2N_1+N_1N_2g^2)m}{(1+(N_1+N_2)g)(3+N_1N_2g^2)-1}\;,
\cr &&\cr
{\tilde \lambda}&=&\frac{2\lambda(1+N_1g)^2+m^2N_1g}{(1+(N_1+N_2)g)
(3+N_1N_2g^2)-1}\;.
\eea
Thus in general one can write $W'(x)=(x-b_1)(x-b_2)$.

\subsection{Unbroken gauge symmetry}

By making use of the fact that this model can be obtained from deformation of
${\cal N}=2$ $U(N)$ SYM theory by adding the superpotential (\ref{WTM}),
the effective superpotential can be obtained from the factorization of 
Seiberg-Witten curve. In fact since the gauge symmetry is not broken
the factorization can be obtained for the confining vacuum where all 
$N-1$ monopoles have condensed in terms of Chebyshev polynomials. Indeed
the solution is the same as (\ref{confsol}). Therefore the effective 
superpotential reads
\be
W_{\rm exact}=N(\lambda+2\Lambda^2+gN\Lambda^2)z+{mN\over 2}(z^2+2\Lambda^2)
+{N\over 3}(1+{3gN\over 2})z^3\;,
\ee
Setting $B:=\Lambda^2$ we can proceed to integrate in the glueball field
$S$ as follows. First we find $B$ in terms of $S$ from the equation
\be
N S=B {\partial W_{\rm exact}\over \partial B}=NB(m+2z+gNz)\;.
\ee
Then we find $z$ by solving
\be
0={\partial W_{\rm exact}\over \partial z}=N
\left((1+{3gN\over 2})z^2+mz+\lambda+2B
+gNB\right)\;.
\ee
The effective action for the glueball superfield $S$ can be written 
as
\be
W_{\rm eff}(S,g,\Lambda)= -S \log\left({B \over \Lambda^2}\right)^N +
W_{\rm exact}(S, g).
\l{mmm}
\ee

To write the effective superpotential explicitly let us, for simplicity, 
set $\lambda=0$. In this case one finds the following solutions
for $z$ and $B$ in power series of $S$ up to order ${\cal O}(S^6)$
\bea
B&=&{S\over m}+\frac{(2+gN)^2S^2}{m^4}+\frac{(10+7gN)(2+gN)^3S^3}{2m^7}\cr &&\cr
&+&\frac{(32+46gN+17g^2N^2)(2+gN)^4S^4}{m^{10}}\cr &&\cr
&+&\frac{(1848+4044gN+3018g^2N^2+769g^3N^3)(2+gN)^5S^5}{8 m^{13}}\cr &&\cr
z&=&-\frac{(2+gN)^2S^2}{m^2}-\frac{(6+5gN)(2+gN)^2S^2}{2m^5}\\ &&\cr &-&
\frac{(16+16gN+11g^2N^2)(2+gN)^3S^3}{m^8}\cr &&\cr
&-&\frac{5(6+5gN)(28+44gN+19g^2N^2)(2+gN)^4S^4}{8m^{11}}\cr &&\cr
&-&\frac{(3072+9768gN+11940g^2N^2+6654g^3N^3+1427g^4N^4)(2+gN)^5S^5}{4 m^{14}}
\nonumber .
\eea
Plugging these solutions into (\ref{mmm}) one gets the effective 
superpotential as follows
\bea
W_{\rm eff}&=&-NS(\log({S\over \Lambda^2})-1)-\frac{N(2+gN)^2S^2}{2m^3}
-\frac{N(4+3gN)(2+gN)^3S^3}{3m^6}\cr &&\cr
&-&\frac{N(140+212gN+83g^2N^2)(2+gN)^4S^4}{24m^{9}}\cr &&\cr
&-&\frac{N(128+292gN+228g^2N^2+61g^3N^3)(2+gN)^5S^5}{4 m^{12}}\;.
\l{WEMUB}
\eea
As a check for this expression we note that setting $g=0$ we will get 
the same result as that in the single trace case.

\subsection{Broken gauge symmetry}

In this case to get a closed form for the effective superpotential we will
consider the case where the gauge symmetry $U(3N)$ is broken down to
$U(2N)\times U(N)$. Essentially this is equivalent to study the case
with $U(3)\rightarrow U(2)\times U(1)$ symmetry breaking. To find the 
effective superpotential one can use the factorization of Seiberg-Witten 
curve as we presented in the appendix.

Regarding the fact that the Seiberg-Witten curve for $U(N)$ theory
is given by
\be
y^2=(x^3-s_1x^2-s_2x-s_3)^2-4\Lambda^6
\ee
basically we need to minimize the total superpotential 
\be
W_{\rm T}=u_3+mu_2+\lambda u_1+gu_1u_2+L\left(p^3-s_1p^2-s_2p-s_3
\pm 2 \Lambda^3\right)
+Q(3p^2-2s_1p-s_2),
\ee
where $p$ could be either $b_1$ or $b_2$. To be specific we set $p=b_1$.
Classically one has
\be
s_1^{\rm class}=2b_1+b_2,\;\;\;\;\;\;\;s_2^{\rm class}=-2b_1b_2+b_1^2,\;\;\;\;
\;\;\;s_3^{\rm class}=b_1^2b_2
\ee
while from the total superpotential, quantum mechanically we find
\be
s_{i}=s_{i}^{\rm class}\pm 2\Lambda^3 \delta_{i,3}\;.
\ee
Therefore the effective superpotential reads
\be
W_{\rm eff}=u_3^{\rm class}+mu_2^{\rm class}+
\lambda u_1^{\rm class}+gu_1^{\rm class}u_2^{\rm class}\pm 2\Lambda^3\;,
\l{pppp}
\ee
where 
\be
u_1^{\rm class}=2b_1+b_2,\;\;\;\;\;u_2^{\rm class}=b_1^2+{1\over 2}b_2^2,\;
\;\;\;\;u_3^{\rm class}={2\over 3} b_1^2+{\over 3}b_2^2\;.
\ee

\section{Linearized superpotential}

\subsection{Field theory description}

Following \cite{Balasubramanian:2002tm} one can recast the superpotential
to the form with only single trace operators using auxiliary fields. In our
case we need two fields $A_1$ and $A_2$ and the superpotential may be written
as
\be
W_{\rm tree}={1\over 3}{\rm Tr}(\phi^3)+{1\over 2}(m+gA_1){\rm Tr}
(\phi^2)+(\lambda+gA_2){\rm Tr}(\phi)-gA_1A_2\;.
\ee
Since $A_1$ and $A_2$ have no dynamics, one can integrate them out and 
refined the multi trace superpotential (\ref{WTM}). These fields can be
treated as constant background fields and therefore the theory can be solved
using single trace superpotential. This will generate an effective
superpotential
$W_{\rm eff}^{\rm single}(A_1,A_2,S)$ as a function of $A_i$'s and glueball
superfield $S$. This function is the same as that in the model without
multi trace but with $A_i$'s dependent couplings. 

For example in the case where the gauge group is not broken the effective 
superpotential can be read from the single trace result (\ref{WTE}) 
\be
W_{\rm exact}^{\rm single}(A_1,A_2)=\langle u_3\rangle +m' \langle
u_2\rangle +\lambda' 
\langle u_1\rangle\;,
\ee
where $m'=m+gA_1,\;\lambda'=\lambda+gA_2$ and
\be
\langle u_1\rangle=Nz,\;\;\;\;\;\langle u_2\rangle={N\over 2}(z^2+2\Lambda^2),
\;\;\;\;\;\langle u_3\rangle={N\over 3}(z^3+6\Lambda^2z)\;.
\ee
The same as in the previous section one can proceed to ``integrate in'' 
the glueball superfield. To do this one sets $B := \Lambda^2$, and uses 
the equation 
\be
N S=B {\partial W_{\rm exact}^{\rm single}\over \partial B}=NB(m'+2z)\;,
\l{SBM}
\ee
to solve for $B$ in terms of $S$. One can also find $z$ by solving
\be
0={\partial W_{\rm exact}^{\rm single}\over \partial
z}=N(z^2+m'z+\lambda'+2B)\;.
\l{ZM}
\ee
Now the effective action for the glueball superfield $S$ can be written 
as
\be
W_{\rm eff}^{\rm single}(A_1,A_2,S)= -S \log\left({B \over
\Lambda^2}\right)^N +
W_{\rm exact}^{\rm single}(A_1,A_2, S)\;.
\ee
Using the result of single trace we get
\bea
W_{\rm eff}^{\rm single}(A_1,A_2,S)&=&
-NS(\log({S\over \Delta' \Lambda^2})-1)\cr &&\cr
&-&{2N\over 3}\;{S^2\over {\Delta'}^3}\left(3+16{S\over {\Delta'}^3}+
140{S^2\over {\Delta'}^6}+512{S^3\over {\Delta'}^9}\right)\;,
\eea
where ${\Delta'}^2=(m+gA_1)^2-4(\lambda+gA_2)$. One should also add to this 
the $-gA_1A_2$ term and the final answer for the 
superpotential is
\be
W_{\rm eff}(A_1,A_2,S)=W_{\rm eff}^{\rm single}(A_1,A_2,S)-gA_1A_2.
\l{rrrr}
\ee
To get the final result for the effective superpotential with multi trace 
operator we need to integrate out $A_i$'s using their equations of motion
\bea
{\partial W_{\rm eff}^{\rm single}(A_1,A_2,S)\over \partial A_1}-g
A_2&=&0\;,\cr 
{\partial W_{\rm eff}^{\rm single}(A_1,A_2,S)\over \partial A_2}-g A_1&=&0\;.
\eea
These equation can be solved to find $A_i$'s in terms of glueball superfield and
then plugging back the results into the (\ref{rrrr}) one can obtain the
effective superpotential for the theory with tree level superpotential
(\ref{WTM}). This should reproduce the field theory result of multi trace
superpotential (\ref{WEMUB}). This can be seen as follows. 

Suppose we have been able to solve the equations (\ref{SBM}) and (\ref{ZM}) 
exactly. Therefore we have the exact form of $z$ and $B$ as a function of 
$S$, $A_1$ and $A_2$
\be
B=B(A_1,A_2,S),\;\;\;\;\;\;z=z(A_1,A_2,S)\;.
\ee
plugging these into the effective superpotential (\ref{rrrr}) one gets 
\bea
W_{\rm eff}(A_1,A_2,S)&=&
{N\over 3}(z^3+6\Lambda^2z)+{N\over 2}(m+gA_1)(z^2+2\Lambda^2)+
(\lambda+gA_2)Nz\cr &&\cr
&-&S \log\left({B \over \Lambda^2}\right)^N -gA_1A_2.
\eea
Thus the equations of motion of $A_i$'s read
\bea
{\partial W_{\rm eff}\over \partial A_1}+ 
{\partial B\over \partial A_1}
{\partial W_{\rm eff}\over \partial B}
+{\partial z\over \partial A_1}
{\partial W_{\rm eff}\over \partial z}-gA_2&=&0\;,\cr &&\cr
{\partial W_{\rm eff}\over \partial A_2}+ 
{\partial B\over \partial A_2}
{\partial W_{\rm eff}\over \partial B}
+{\partial z\over \partial A_2}
{\partial W_{\rm eff}\over \partial z}-gA_1&=&0\;,
\eea
which are 
\bea
0&=&{gN\over 2}(z^2+2B)-gA_2+N{\partial B\over \partial A_1}
\left(-{S\over B}+(m+z+gA_1)\right)\cr &&\cr
&+&N{\partial z\over \partial A_1}\left(z^2+(m+gA_1)z+\lambda+gA_2+B
\right)\;,\cr &&\cr
0&=&gNz-gA_1+N{\partial B\over \partial A_2}
\left(-{S\over B}+(m+z+gA_1)\right)\cr &&\cr
&+&N{\partial z\over \partial A_2}\left(z^2+(m+gA_1)z+\lambda+gA_2+B
\right)\;.
\eea
By making use of the equations (\ref{SBM}) and (\ref{ZM}) we find
\be
A_2 ={N\over 2}(z^2+2B),\;\;\;\;\;\;   A_1=Nz\;.
\ee 
Now one has to plug these solution into the effective superpotential 
to get the final result which is, of course, what we have found in the 
previous section (\ref{WEMUB}).

In the case where the gauge group is broken to two parts we can follow
the same procedure. To be specific we consider $U(3)\rightarrow U(2)
\times U(1)$ where we will be able to write a closed form for the
exact superpotential. More precisely using the field theory 
result in the single trace case the effective superpotential reads 
\be
W_{\rm eff}(A_1,A_2)=(\lambda +gA_2){u'}_1^{\rm class} +
(m+gA_1){u'}_2^{\rm class}
+{u'}_3^{\rm class}
\pm 2\Lambda^3-gA_1A_2\;,
\l{ppp}
\ee
where 
\be
{u'}_1^{\rm class}= 2a'_1+a'_2\;,  \;\;\;\;\;\; {u'}_2^{\rm class}=
{a'}_1^2+\frac{{a'}_2^2}{2}\;,
\;\;\;\;\;\;
{u'}_3^{\rm class}= \frac{2{a'}_1^3}{3}+\frac{{a'}_2^3}{3} \pm 2\Lambda^3\;,
\ee
with $a'_{1,2}=-{m'\over 2}\pm {1\over 2}\sqrt{{m'}^2-4\lambda'}$.
We should now show that upon integrating out the auxiliary fields
$A_1$ and $A_2$ the obtained effective action is the same as 
that in the field theory computation with multi trace operator
(\ref{pppp}). To see this we note that
\bea
\frac{\partial W_{\rm eff}}{\partial A_1}= g ({u'}_2^{\rm class}- A_2)
+\left( (\lambda +gA_2)\frac{\partial {u'}_1^{\rm class}}
{\partial A_1} +(m+gA_1)\frac{\partial {u'}_2^{\rm class}}
{\partial A_1}+
\frac{\partial {u'}_3^{\rm class}}{\partial A_1}\right) =0, \cr \cr
\frac{\partial W_{\rm eff}}{\partial A_2}= g ({u'}_1^{\rm class}- A_1)
+\left( (\lambda +gA_2)\frac{\partial {u'}_1^{\rm class}}
{\partial A_2} +(m+gA_1)\frac{\partial {u'}_2^{\rm class}}
{\partial A_2}+
\frac{\partial {u'}_3^{\rm class}}{\partial A_2}\right)=0,
\eea
which leads to the following solution for $A_i$'s
\be
A_1={u'}_1^{\rm class},\;\;\;\;\;A_2={u'}_2^{\rm class}\;.
\ee 
From these expressions one can find $A_{1}$ and $A_2$ and plugging
them into the effective superpotential (\ref{ppp}). Doing so we
will get the same result as (\ref{pppp}).

\subsection{Matrix model description}
	
In this section we study the matrix model of the gauge theory with
multi trace operators. As it was shown in \cite{Balasubramanian:2002tm}
taking the naively $W$ including a multi trace operator as the 
potential of the corresponding matrix model would lead to an incorrect
result. And in fact we should work with the linearized form of the
superpotential. Therefore we consider $U(M)$ matrix model with 
the following cubic potential
\be
W_{\rm tree}={1\over 3}\;{\rm Tr}(\phi^3)+{1\over 2}\;m'\;
{\rm Tr}(\phi^2)+\lambda'\;
{\rm Tr}(\phi)-gA_1A_2\;.
\ee
This can be thought of as a matrix model with the single trace potential
while treating $A_i$'s as constant background fields plus a shift of the
form $-gA_1A_2$. For the single trace part, the potential has two critical 
point $a_1'$ and $a_2'$ such that
\be
W'(x)=(x-a'_1)(x-a'_2),\;\;\;\;\;a'_{1,2}=-{m'\over 2}\pm
{1\over 2}\sqrt{m'^2-4\lambda'}\;.
\ee
In the case where the gauge symmetry is not broken one can take the
following small fluctuations 
\be
\phi=a'_1 1_{M\times M}+\varphi\;,
\ee
and therefore the potential of matrix model reads
\be
W(\varphi)=W(a'_1)+{1\over 3}\;{\rm Tr}(\varphi^3)+{1\over 2}\;\Delta'\;
{\rm Tr}(\varphi^2)\;,
\ee
where $\Delta=a'_1-a'_2$. We can now write down the Feynman rules and 
thereby evaluate the free energy order by order. Here we shall also
consider the large $M$ limit while keeping $g_sM=S$ fixed. Thus
only planar diagram would contribute. Basically using the single trace result 
as that in section 2 we find
\be
{\cal F}_0^{\rm single}
(A_1,A_2,S)=-{1\over 2}S^2\log\left({S\over {\Delta'}^3}\right)+S^2
\log\left({\Lambda\over \Delta'}\right)+{2\over 3}\;{S^3\over{\Delta'}^3}
\left(1+4{S\over {\Delta'}^3}+28{S^2\over {\Delta'}^6}\right)
\ee
up to 4-loop. Using this expression, the exact superpotential is given by
\be
W_{\rm eff}^{\rm single}(A_1,A_2,S)=-NS\left(\log\left({S\over
{\Delta'}\Lambda^2}\right)-1\right)
-{2N\over 3}\;{S^2\over {\Delta'}^3}
\left(3+16{S\over {\Delta'}^3}+140{S^2\over {\Delta'}^6}\right)
\ee
Finally the effective superpotential for the multi trace model can be
found by integrating out $A_1$ and $A_2$ from the total superpotential
given by
\be
W_{\rm eff}(A_1,A_2,S)=W_{\rm eff}^{\rm single}(A_1,A_2,S)-gA_1A_2\;,
\ee
This, of course, is the same expression as (\ref{rrrr}) and thus would lead
to correct answer. Therefore we might conclude that the linearized 
superpotential would give a correct matrix model for an ${\cal N}=1$ gauge 
theory with a multi trace operators in the superpotential.

On the other hand for the case where the gauge group is broken, 
we consider the large $M$ $U(M)$ matrix model and take the small fluctuations 
as follows
\be
\phi=\pmatrix{a_11_{M_1\times M_1} & 0\cr 0 & a_21_{M_2\times M_2}}+
\pmatrix{\varphi_{11}&\varphi_{12}\cr \varphi_{21}&\varphi_{22}}\;,
\ee 
with $M_1+M_2=M$. Therefore the gauge symmetry is broken down to 
$U(M_1)\times U(M_2)$. We shall also consider the large $M_1$ and $M_2$ 
limit while keeping $S_1=g_sM_1$ and $S_2=g_sM_2$ fixed. Using the single
trace result the matrix model action is found to be
\bea
W&=&{1 \over 2}\;\Delta'\bigg{(}{\rm Tr}(\varphi_{11}^2)-{\rm
Tr}(\varphi_{22}^2)
\bigg{)}+{1\over 3}\bigg{(}{\rm Tr}(\varphi_{11}^3)+{\rm Tr}(\varphi_{22}^3)
\bigg{)}\cr &&\cr
&+&\Delta'\bigg{(}{\rm Tr}(B_{21}C_{12})-{\rm Tr}(B_{12}C_{21})\bigg{)}+
{\rm Tr}(B_{21}\varphi_{11}C_{12}+C_{21}\varphi_{11}B_{12})\cr &&\cr
&+&{\rm Tr}(B_{12}\varphi_{22}C_{21}+C_{11}\varphi_{22}B_{21})\;.
\eea
Correspondingly the matrix model free energy up to 4-loop reads 
\bea
&&{\cal F}_{0}(A_1,A_2,S_1,S_2)=-{1\over 2}\sum
{S_i}^2\log\left({S_i\over {\Delta'}^3}
\right)+(S_1+S_2)^2 \log\left({\Lambda\over {\Delta'}^3}\right)\cr &&\cr
&&\;\;\;\;+{1\over 3{\Delta'}^3}
\bigg{(}2S_1^3-15S_1^2S_2+15S_1S_2^2-2S_2^3\bigg{)}\cr &&\cr
&&\;\;\;\;+{1\over 3{\Delta'}^6}
\bigg{(}8S_1^4-91S_1^3S_2+177S_1^2S_2^2-91S_1S_2^3-8S_2^4\bigg{)}
\cr &&\cr
&&\;\;\;\;+{1\over 3{\Delta'}^9}
\bigg{(}56S_1^5-871S_1^4S_2+2636S_1^3S_2^2-2636S_1^2S_2^3
+871S_1S_2^4-56S_2^5\bigg{)}.
\eea
Having the explicit expression for the matrix model free energy with 
symmetry breaking as $U(M)\rightarrow U(M_1)\times U(M_2)$ one can 
find the effective superpotential $W_{\rm eff}^{\rm single}(A_i,S_i)$
for the gauge theory where the gauge group is broken as 
$U(N)\rightarrow U(N_1)\times U(N_2)$ by making
use of (\ref{lll}). Then the effective superpotential for the multi
trace theory can be obtained by integrating out the auxiliary fields
$A_i$'s from 
\be
W_{\rm eff}(A_i,S_i)=W_{\rm eff}^{\rm single}(A_i,S_i)-gA_1A_2\;.
\ee

To check the result one might consider the model with $N_1=2$ and 
$N_2=1$ where the field theory result is known. Doing the same 
analysis as before one can see that this does give the correct 
answer.

\section{Conclusions}

In this paper we have studied ${\cal N}=1$ supersymmetric 
$U(N)$ gauge theory coupled to an adjoint scalar superfield 
with a cubic superpotential containing a multi trace 
operator. Then we have looked for the corresponding matrix model
in the context of the Dijkgraaf-Vafa's proposal 

Following \cite{Balasubramanian:2002tm} we have considered a matrix model
in which its potential is given by linearized form of the superpotential
of the corresponding gauge theory using some auxiliary fields. 
In this way the problem can be recast to the single trace case with, of 
course, coefficients which now depend on the auxiliary fields. Using
this matrix model one can find the free energy and thereby the effective
superpotential using the Dijkgraaf-Vafa's proposal. At the end we should
integrate out the auxiliary fields finding the final result of the
exact superpotential for the theory with multi trace in the tree level
superpotential. As it was noticed in \cite{Balasubramanian:2002tm} it is
crucial when the auxiliary fields are integrated out.

In this paper we have only considered the multi trace operator with 
the form ${\rm Tr}(\phi){\rm Tr}(\phi^2)$, while we could have also
considered other multi trace operators like 
$\left({\rm Tr}(\phi)\right)^3$. In this 
paper we have studied two different models: one with gauge symmetry
breaking and the other without gauge symmetry breaking. In both cases
we have seen that linearized matrix model does give the 
correct field theory result. 

In fact one of our motivation for doing this project was whether the 
Dijkgraaf-Vafa's proposal can be also applied for the exceptional
group. We note, however, that the tree level superpotential
of a gauge theory with an exceptional group has usually multi trace
operators. For example ${\cal N}=1$ supersymmetric gauge theory with 
gauge group $G_2$ can be obtained from ${\cal N}=2$ 
$G_2$ SYM theory by a tree level superpotential given by 
\be
W_{\rm tree}={m\over 4}{\rm Tr}(\phi^2)+
{g\over 6}\left({\rm Tr}(\phi^6)-{1\over 16}{\rm Tr}(\phi^2)^3\right).
\ee
So the first step to study these theories is to increase our 
knowledge about the physics of multi trace operators. We hope
to address this issue in the future.

\vspace*{.4cm}

{ \bf Acknowledgements}

We would like to thank Hessam Arfaei and Cumrun Vafa for useful discussions 
and comments. We would also like to thank Vijay Balasubramanian, 
Mohammad R. Garousi and Shahrokh Parvizi for comments.

\appendix

\section{Appendix}

In this appendix we show how the factorization of Seiberg-Witten curve
can be worked out for the case where the gauge symmetry $U(N)$ is broken
down to $U(2)\times U(1)^{N-1}$. To do this Consider $\cn =2$ SYM theory with  
$U(N)$ gauge group. The corresponding
Seiberg-Witten curve is given by \cite{{Argyres:1994xh},{Klemm:1995wp}}
\be
y^2=P(x,s_i)^2-4\Lambda^{2N}\;,\;\;\;\;\;{\rm with}\;\;
P(x,s_i)=x^N-\sum_{i=1}^Ns_{i}x^{N-i}\;,
\l{SUCUR}
\ee
where
\be
ks_k+\sum_{i=1}^ki s_{k-i}u_i=0\;,\;\;\;\;\;k=1,2,\cdots, N,
\ee
with $s_0=-1$. Here $u_k$'s are the gauge invariant 
parameters of the theory defined in terms of $\cn=1$ adjoint superfiled
$\phi$ as following
\be
u_k={1\over k}{\rm Tr}(\phi^k)
\ee

We would like to perturb the theory by adding the following a tree-level 
superpotential
\be
W=\sum_{n=1}^N{g_n\over n} {\rm Tr}(\phi^n)\;.
\l{SuperSU}
\ee
The supersymmetric $\cn=1$ vacua of the theory are determined by
F-term condition $W'=\sum_ng_n\phi^{n-1}=0$. Then the roots $a_i$
of  
\be
W'(x)=\sum_{n=1}^{N}g_n x^{n-1}=g_N\prod_{i=1}^{N-1}(x-a_i)
\ee
give the eigenvalues of $\phi$. In particular we are interested in the
vacuum where the gauge group $U(N)$ is broken down to $U(2)\times
U(1)^{N-1}$ in which in low energies we are left with an $\cn=1$
$SU(2)$ SYM theory which is in confining phase and the photon
multiplets for $U(1)^{N-1}$ are decoupled.  Therefore we take
\be
\phi={\rm diag}(a_1,a_1,a_2,a_3,\cdots,a_{N-1})\;.
\ee
 
This $\cn=1$ vacuum where one monopole becomes massless is
parameterized by the set of moduli ${{\tilde s}_i}$ where the 
Seiberg-Witten curve factorizes in such a way that (\ref{SUCUR}) has one 
double root and $2(N-1)$ single roots:
\be
P(x,{\tilde s}_i)^2-4\Lambda^{2N}=H_1^2(x)F_{2(N-1)}(x)\;.
\l{Factor}
\ee
where $H_1$ is given by $H_1(x)=x-x_0$ with some $x_0$ to be determined.
Therefore the factorization (\ref{Factor}) is equivalent to
\be
P(x_0,{\tilde s}_i)\pm 2\Lambda^N=0\;,\;\;\;\;\;{\partial 
P(x,{\tilde s}_i)\over \partial x}|_{x=x_0}=0\;.
\ee

Now the taste is to minimize the superpotential (\ref{SuperSU}) subject
to the above constraints. Thus the total superpotential can be written as
\be
W_{\rm T}=\sum_{n=1}^{N}g_nu_n+L\left(P(x_0,{\tilde s}_i)\pm 2 \Lambda^N\right)
+Q\;P'(x_0,{\tilde s}_i)\;.
\ee
Here $L,Q$ and $x_0$ should be treated as Lagrange multipliers. From the total
superpotential the equations of motion for $L,Q$ and $x_0$ read
\bea
{\partial\over \partial L}&:& P(x_0,{\tilde s}_i)\pm 2\Lambda^N=0\;,\cr
&&\cr
{\partial\over \partial Q}&:& P'(x_0,{\tilde s}_i)=0\;,\cr &&\cr
{\partial\over \partial x_0}&:& Q P''(x_0,{\tilde s}_i)=0\;.
\l{eq1SU}
\eea
Moreover the equation of motion for $u_k$ leads to
\be
g_k=L{\partial\over \partial u_{k}}P(x_0,{\tilde
s}_i)=-L\sum_{i=2}^{N}x_0^{N-i}
{\partial {\tilde s}_i \over \partial u_{k}}=-L\sum_{i=2}^{N}x_0^{N-i}
{\tilde s}_{i-k}\;.
\l{eq2SU}
\ee
Here in the last equality we have used the Newton's relation to get
${\partial {\tilde s}_i \over \partial u_{k}}={\tilde s}_{i-k}$.
The equations (\ref{eq1SU}) and (\ref{eq2SU}) can be solved for the parameters. 
Doing so
one finds
\be
Q=0,\;\;\;\;L=g_N,\;\;\;x_0=a_1,\;\;\;\;{\tilde s}_i=s_i\pm
2\Lambda^N\delta_{iN}\;.
\ee
Plugging these solutions into the total superpotential one gets
\be
W_{\rm eff}=\sum_{k=1}^Ng_ku_k\pm 2g_N\Lambda^N\;.
\l{exP}
\ee
This is the same expression obtained in \cite{Terashima:1996pn}.
It can also be show that
\bea
F_{2(N-1)}&=&\prod_{i=1}^{N-1}(x-a_i)^2\mp 4\Lambda^N 
\prod_{i=2}^{N-1}(x-a_i)\cr &&\cr
&=&{1\over g_N^2}\left({W'(x)}^2+f_{N-2}\right)
\eea
with $f_{N-2}=\mp 4g_N^2\Lambda^N \prod_{i=2}^{N-1}(x-a_i)$ being the quantum
correction. This means that the quantum dynamics of the $\cn=1\;U(1)^{N-2}$ at
low energies is captured by the following curve 
\be
y^2={W'(x)}^2+f_{N-2}\;.
\l{recSU}
\ee
Having the reduced curve explicitly one can proceed to evaluate the periods
of the curve. The periods are given in terms of integral of a one form over
different one cycles of the curve
\be
S_i=\oint_{\alpha_i}ydx=\oint_{\alpha_i}dx\; g_N\sqrt{
\prod_{i=1}^{N-1}(x-a_i)^2
\mp 4\Lambda^N 
\prod_{i=2}^{N-1}(x-a_i)}\;,
\ee
where $\alpha_i$ is a one cycle loops around the $i$th cut of the reduced
curve. In particular one has
\be
S=\sum_{i}S_i=\oint_{C}dx\; g_N\sqrt{
\prod_{i=1}^{N-1}(x-a_i)^2
\mp 4\Lambda^N 
\prod_{i=2}^{N-1}(x-a_i)}\;,
\ee
where $C$ is a loop at infinity. Therefore we get
\bea
S&=&\oint_{C}dx\; g_N\prod_{i=1}^{N-1}(x-a_i)\left(
1\mp {2\Lambda^N \over (x-a_1)\prod_{i=1}^{N-1}(x-a_i)}+\cdots\right)
\cr &&\cr
&=&\pm 2 g_N\Lambda^N\;,
\eea
which means
\be
{\partial W_{\rm eff}\over \partial(\log\Lambda^{2N})}=-{b_{N-2}\over
4g_N}\;,
\ee
where $b_{N-2}=\mp 4g_N^2\Lambda^N$ is the numerical coefficient of $x^{N-2}$
term in the $f_{N-2}$. In other words we get
\be
\Lambda^2{\partial W_{\rm eff}\over \partial \Lambda^2}=NS\;,
\ee
which can be interpreted as the Konishi anomaly 
\cite{{Konishi:1983hf},{Konishi:1985tu},{Cachazo:2002ry}}.


\begin{thebibliography}{99}

\bibitem{Dijkgraaf:2002dh}
R.~Dijkgraaf and C.~Vafa,
``A perturbative window into non-perturbative physics,''
arXiv:hep-th/0208048.

\bibitem{Bershadsky:1993cx}
M.~Bershadsky, S.~Cecotti, H.~Ooguri and C.~Vafa,
``Kodaira-Spencer theory of gravity and exact results for quantum 
string amplitudes,''
Commun.\ Math.\ Phys.\  {\bf 165}, 311 (1994)
[arXiv:hep-th/9309140].

\bibitem{Cachazo:2001jy}
F.~Cachazo, K.~A.~Intriligator and C.~Vafa,
``A large N duality via a geometric transition,''
Nucl.\ Phys.\ B {\bf 603}, 3 (2001)
[arXiv:hep-th/0103067].

\bibitem{Cachazo:2002pr}
F.~Cachazo and C.~Vafa,
``${\cal N}=1$  and ${\cal N}=2$ geometry from fluxes,''
arXiv:hep-th/0206017.

\bibitem{Dijkgraaf:2002fc}
R.~Dijkgraaf and C.~Vafa,
``Matrix models, topological strings, and supersymmetric gauge theories,''
Nucl.\ Phys.\ B {\bf 644}, 3 (2002)
[arXiv:hep-th/0206255].

\bibitem{Dijkgraaf:2002vw}
R.~Dijkgraaf and C.~Vafa,
``On geometry and matrix models,''
Nucl.\ Phys.\ B {\bf 644}, 21 (2002)
[arXiv:hep-th/0207106].

\bibitem{Dijkgraaf:2002xd}
R.~Dijkgraaf, M.~T.~Grisaru, C.~S.~Lam, C.~Vafa and D.~Zanon,
``Perturbative computation of glueball superpotentials,''
arXiv:hep-th/0211017.

\bibitem{Gorsky:2002uk}
A.~Gorsky,
``Konishi anomaly and ${\cal N}=1$ effective superpotentials from 
matrix models,''
Phys.\ Lett.\ B {\bf 554}, 185 (2003)
[arXiv:hep-th/0210281].


\bibitem{Cachazo:2002ry}
F.~Cachazo, M.~R.~Douglas, N.~Seiberg and E.~Witten,
``Chiral rings and anomalies in supersymmetric gauge theory,''
JHEP {\bf 0212}, 071 (2002)
[arXiv:hep-th/0211170].

\bibitem{Seiberg:2002jq}
N.~Seiberg,
``Adding fundamental matter to 'Chiral rings and anomalies in  
supersymmetric gauge theory',''
JHEP {\bf 0301}, 061 (2003)
[arXiv:hep-th/0212225].

\bibitem{Tachikawa:2002wk}
Y.~Tachikawa,
``Derivation of the Konishi anomaly relation from Dijkgraaf-Vafa 
with  (bi-)fundamental matters,''
arXiv:hep-th/0211189.


\bibitem{Cachazo:2002zk}
F.~Cachazo, N.~Seiberg and E.~Witten,
``Phases of ${\cal N}=1$ supersymmetric gauge theories 
and matrices,''
JHEP {\bf 0302}, 042 (2003)
[arXiv:hep-th/0301006].

\bibitem{Roiban:2003uq}
R.~Roiban, R.~Tatar and J.~Walcher,
``Massless flavor in geometry and matrix models,''
arXiv:hep-th/0301217.

\bibitem{Brandhuber:2003va}
A.~Brandhuber, H.~Ita, H.~Nieder, Y.~Oz and C.~Romelsberger,
``Chiral Rings, Superpotentials and the Vacuum 
Structure of ${\cal N}=1$ Supersymmetric Gauge Theories,''
arXiv:hep-th/0303001.



\bibitem{Balasubramanian:2002tm}
V.~Balasubramanian, J.~de Boer, B.~Feng, Y.~H.~He,
 M.~x.~Huang, V.~Jejjala and A.~Naqvi,
``Multi-trace superpotentials vs. Matrix models,''
arXiv:hep-th/0212082.

\bibitem{Dijkgraaf:2002pp}
R.~Dijkgraaf, S.~Gukov, V.~A.~Kazakov and C.~Vafa,
``Perturbative analysis of gauged matrix models,''
arXiv:hep-th/0210238.

\bibitem{Ferrari:2002jp}
F.~Ferrari,
``On exact superpotentials in confining vacua,''
Nucl.\ Phys.\ B {\bf 648}, 161 (2003)
[arXiv:hep-th/0210135].

\bibitem{Ferrari:2002kq}
F.~Ferrari,
``Quantum parameter space and double scaling limits in 
${\cal N}=1$ super  Yang-Mills theory,''
arXiv:hep-th/0211069.

\bibitem{Klemm:2002pa}
A.~Klemm, M.~Marino and S.~Theisen,
``Gravitational corrections in supersymmetric gauge theory 
and matrix  models,''
arXiv:hep-th/0211216.

\bibitem{Dijkgraaf:2002yn}
R.~Dijkgraaf, A.~Sinkovics and M.~Temurhan,
``Matrix models and gravitational corrections,''
arXiv:hep-th/0211241.


\bibitem{Douglas:1995nw}
M.~R.~Douglas and S.~H.~Shenker,
``Dynamics of $SU(N)$ supersymmetric gauge theory,''
Nucl.\ Phys.\ B {\bf 447}, 271 (1995)
[arXiv:hep-th/9503163].

\bibitem{Intriligator:1994uk}
K.~A.~Intriligator,
``'Integrating in' and exact superpotentials in 4-d,''
Phys.\ Lett.\ B {\bf 336}, 409 (1994)
[arXiv:hep-th/9407106].


\bibitem{Elitzur:1996gk}
S.~Elitzur, A.~Forge, A.~Giveon, K.~A.~Intriligator and E.~Rabinovici,
``Massless Monopoles Via Confining Phase Superpotentials,''
Phys.\ Lett.\ B {\bf 379}, 121 (1996)
[arXiv:hep-th/9603051].


\bibitem{Argyres:1994xh}
P.~C.~Argyres and A.~E.~Faraggi,
``The vacuum structure and spectrum of $\cn=2$ supersymmetric $SU(N)$
gauge theory,''
Phys.\ Rev.\ Lett.\  {\bf 74}, 3931 (1995)
[arXiv:hep-th/9411057].

\bibitem{Klemm:1995wp}
A.~Klemm, W.~Lerche, S.~Yankielowicz and S.~Theisen,
``On the monodromies of $\cn=2$ supersymmetric Yang-Mills theory,''
arXiv:hep-th/9412158.

A.~Klemm, W.~Lerche and S.~Theisen,
``Nonperturbative effective actions of $\cn=2$ supersymmetric 
gauge theories,''
Int.\ J.\ Mod.\ Phys.\ A {\bf 11}, 1929 (1996)
[arXiv:hep-th/9505150].

\bibitem{Terashima:1996pn}
S.~Terashima and S.~K.~Yang,
``Confining phase of $\cn=1$ supersymmetric gauge theories and $\cn= 2$
  massless solitons,''
Phys.\ Lett.\ B {\bf 391}, 107 (1997)
[arXiv:hep-th/9607151].

\bibitem{Konishi:1983hf}
K.~Konishi,
``Anomalous Supersymmetry Transformation Of Some Composite 
Operators In SQCD,''
Phys.\ Lett.\ B {\bf 135}, 439 (1984).

\bibitem{Konishi:1985tu}
K.~i.~Konishi and K.~i.~Shizuya,
``Functional Integral Approach To Chiral Anomalies In Supersymmetric 
Gauge Theories,''
Nuovo Cim.\ A {\bf 90}, 111 (1985).



\end{thebibliography}
\end{document}